\input harvmac
\input epsf

%%%%%%%%%%%%%%%%%%%%%%%%%%%%%%%%%%%%%%%%%%%%%%%%%%%%%%%%%%%%%%%%%%%%%%
%%%%%%%%%%%%%%%%%%%%%%%        references         
%%%%%%%%%%%%%%%%%%%%%%%%%%%%%%%%%%%%%%%%%%%%%%%%%%%%%%%%%%%%%%%%%%%%%%
\lref\rNojiri{S. Nojiri and S. Odintsov, {\it Two-Boundaries AdS/CFT Correspondence 
in Dilatonic Gravity,} Phys.Lett. {bf B449} (1999) 39-47, {\tt hep-th/9812017}.}
%%CITATION=PHLTA,B449,39;%%
\lref\rAhn{C. Ahn, K. Oh and R. Tatar, {\it The Large N Limit of ${\cal N}=1$ 
Field Theories from F Theory,} Mod.Phys.Lett. {bf A14} (1999) 369-378, {\tt hep-th/9808143}.} 
%%CITATION=MPLAE,A14,369;%%
\lref\rHyun{S. Hyun, Y. Kiem and H. Shin, {\it Effective Action for Membrane Dynamics 
in DLCQ $M$ theory on a Two-torus,} Phys. Rev. {\bf D59} (1999) 021901, {\tt hep-th/9808183}.}
%%CITATION=PHRVA,D59,021901;%%
\lref\rSW{N. Seiberg and E. Witten, {\it Monopoles, Duality and Chiral
Symmetry Breaking in N=2 Supersymmetric QCD,} Nucl. Phys. {\bf B431}
(1994) 484, {\tt hep-th/9408099.}}
%%CITATION=NUPHA,B431,484;%%
\lref\rdorey{N. Dorey, V.V. Khoze and M.P. Mattis, {\it On N=2 
Supersymmetric QCD with Four Flavors,} Nucl.Phys. {\bf B492} (1997) 
607, {\tt hep-th/9611016}.}
%%CITATION=NUPHA,B492,607;%%
\lref\rHen{M. Henningson, {\it Extended superspace, higher derivatives and 
SL(2,Z) duality,} Nucl. Phys. {\bf B458} (1996) 445, {\tt hep-th/9507135}.}
%%CITATION=NUPHA,B458,445;%%
\lref\rDS{M. Dine and N. Seiberg, 
{\it Comments on Higher Derivative Operators in
some SUSY Field Theories,} Phys.Lett. {\bf B409} (1997) 239, 
{\tt hep-th/9705057}.}
%%CITATION=PHLTA,B409,239;%%
\lref\rKD{N. Dorey, V.V. Khoze, M.P. Mattis, M.J. Slater and W.A. Weir,
{\it Instantons, Higher Derivative Terms and Non-renormalization Theorems in
Supersymmetric Gauge Theories,} Phys.Lett. {\bf B408} (1997) 213, 
{\tt hep-th/9706007}.}
%%CITATION=PHLTA,B408,213;%%
\lref\rOvrut{I.L. Buchbinder, S.M. Kuzenko and B.A. Ovrut,
{\it On the D=4 N=2 Nonrenormalization Theorem,} Phys. Lett. {\bf B433}
(1998) 335, {\tt hep-th/9710142.}}
%%CITATION=PHLTA,B433,335;%%
\lref\rOLoop{B. de Wit, M.T. Grisaru and M. Rocek, {\it Non-holomorphic
Corrections to the One Loop N=2 Super Yang-Mills Action,}
Phys. Lett. {\bf B374} (1996) 297, {\tt hep-th/9601115}
%%CITATION=PHLTA,B374,297;%%
\semi
F. Gonzalez-Rey, U. Lindstrom, M. Rocek and R. Von Unge, {\it On N=2
Low Energy Efective Actions,} Phys. Lett. {\bf B388} (1996) 581, 
{\tt hep-th/9607089}
%%CITATION=PHLTA,B388,581;%%
\semi
S. Ketov {\it On the next-to-leading-order correction to the effective action
in N=2 gauge theories,} Phys. Rev. {\bf D57} (1998) 1277,
{\tt hep-th/9706079.}}
%%CITATION=PHRVA,D57,1277;%%
\lref\rocek{F. Gonzalez-Rey, B. Kulik, I.Y. Park and M. Rocek, {\it Selfdual 
Effective Action of N=4 super Yang-Mills,} {\tt hep-th/9810152}.}
%%CITATION=HEP-TH 9810152;%%
\lref\rOfer{O. Aharony, A. Fayyazuddin and J. Maldacena, {\it The Large N
Limit of N=2,1 Field Theories from Threebranes in F-theory,} JHEP 9807
(1998) 013, {\tt hep-th/9806159}.}
%%CITATION=HEP-TH 9806159;%%
\lref\rGSVY{B.R. Greene, A.Shapere, C.Vafa and S.T.Yau, {\it Stringy Cosmic
Strings and Noncompact Calabi-Yau Manifolds,} Nucl. Phys. {\bf B337}
(1990) 1
%%CITATION=NUPHA,B337,1;%%
\semi
M. Asano, {\it Stringy Cosmic Strings and Compactifications of F Theory,}
Nucl. Phys. {\bf B503} (1997) 177,  {\tt hep-th/9703070}.}
%%CITATION=NUPHA,B503,177;%
\lref\rKeh{A. Kehagias, {\it New Type IIB vacua and their F-Theory
Interpretation,} Phys. Lett. {\bf B435} (1998) 337, 
{\tt hep-th/9805131}.}
%%CITATION=PHLTA,B435,337;%%
\lref\rDoug{A. Sen, {\it F Theory and Orientifolds,} Nucl. Phys.
{\bf B475} (1996) 562, {\tt hep-th/9605150}
%%CITATION=NUPHA,B475,562;%%
\semi
T.Banks, M.R.Douglas and N.Seiberg, {\it Probing F Theory with branes,}
Phys. Lett. {\bf B387} (1996) 74, {\tt hep-th/9605199}
%%CITATION=PHLTA,B387,74;%%
\semi
M.R.Douglas, D.A.Lowe and J.H.Schwarz, {\it Probing F Theory with multiple
branes,} Phys. Lett. {\bf B394} (1997) 297, {\tt hep-th/9612062}.}
%%CITATION=PHLTA,B394,297;%%
\lref\rJev{A. Jevicki and T. Yoneya, {\it Spacetime Uncertainty Principle
and Conformal Symmetry in D-particle Dynamics,} Nucl. Phys. {\bf B535}
(1998) 335, {\tt hep-th/9805069}
%%CITATION=NUPHA,B535,335;%%
\semi
A. Jevicki, Y. Kazama and T. Yoneya, {\it Quantum Metamorphosis
of Conformal Transformation in D3 Brane Yang-Mills Theory,}
Phys. Rev. Lett. {\bf 81} (1998) 5072, {\tt hep-th/9808039}
%%CITATION=PRLTA,81,5072;%%
\semi
A. Jevicki, Y. Kazama and T. Yoneya, {\it Generalized Conformal Symmetry
in D Brane Matrix Models,} Phys. Rev. {\bf D59} (1999) 066001,
{\tt hep-th/9810146}}
%%CITATION=PHRVA,D59,066001;%%
\lref\rHo{V. Periwall and R. von Unge, {\it Accelerating D Branes,} Phys. 
Lett. {\bf B430} (1998) 71, {\tt hep-th/9801121}
%%CITATION=PHLTA,B430,71%%
\semi
J. de Boer, K. Hori, H. Ooguri and Y. Oz, {\it Kahler Potential and Higher 
Derivative Terms from M Theory Fivebranes,} Nucl. Phys. B518 (1998) 173-211,
{\tt hep-th/9711143}. }
%%CITATION=NUPHA,B518,173;%%
\lref\rKh{N. Dorey, V.V. Khoze and M.P. Mattis, {\it Multi Instanton Calculus 
in N=2 Supersymmetric Gauge Theory. 2. Coupling to Matter} Phys. Rev. 
{\bf D54} (1996) 7832, {\tt hep-th/9607202}.}
%%CITATION=PHRVA,D54,7832;%%
\lref\rDaction{A.A. Tseytlin, {\it Self-duality of Born-Infeld action and Dirichlet
 3-brane of type IIB superstring theory,} Nucl. Phys. {\bf B469} (1996) 51, 
{\tt hep-th/9602064}
%%CITATION=NUPHA,B469,51;%%
\semi
M. Aganic, J. Park, C. Popescu and J. Schwarz, {\it Dual D-Brane Actions,}
Nucl. Phys. {\bf B496} (1997) 215, {\tt hep-th/9702133.}}
%%CITATION=NUPHA,B496,215;%%
\lref\rTseyt{I. Chepelev and A. Tseytlin, {\it On Membrane Interaction in
Matrix Theory,} Nucl. Phys. {\bf B524} (1998) 69, {\tt hep-th/9801120.}}
%%CITATION=NUPHA,B524,69;%%
\lref\rBob{R. de Mello Koch and R. Tatar, {\it Higher Derivative Terms
from Threebranes in F Theory,} Phys.Lett. {\bf B450} (1999) 99, 
{\tt hep-th/9811128}.}
%%CITATION=PHLTA,B450,99;%%
\lref\rKD{A. Yung, {\it Instanton Induced Effective Lagrangian in the 
Seiberg-Witten Model,} Nucl. Phys. {\bf B485} (1997) 38, 
{\tt hep-th/9605096}
%%CITATION=NUPHA,B485,38;%%
\semi
N. Dorey, V.V. Khoze, M.P. Mattis, M.J. Slater and W.A. Weir,
{\it Instantons, Higher Derivative Terms and Non-renormalization Theorems in
Supersymmetric Gauge Theories,} Phys.Lett. {\bf B408} (1997) 213, 
{\tt hep-th/9706007}
%%CITATION=PHLTA,B408,213;%%
\semi
D. Bellisai, F. Fucito, M. Matone and G. Travaglini, {\it Nonholomorphic terms
in N=2 Susy Wilsonian Actions and RG Equation,} Phys. Rev. {\bf D56}
(1997) 5218, {\tt hep-th/9706099.}}
%%CITATION=PHRVA,D56,5218;%%
\lref\rDas{S.R. Das, {\it Brane Waves, Yang-Mills Theory and Causality,}
{\tt hep-th/9901006.}}
%%CITATION=HEP-TH 9901006;%%
\lref\rAlv{E. Alvarez and C. Gomez, {\it Noncritical Coinfining Strings and 
the Renormalization Group,} {\tt hep-th/9902012.}}
%%CITATION=HEP-TH 9902012;%%
\lref\rDie{I. Antoniadis, {\it A Possible New Dimension at a Few TeV,}
Phys. Lett. {\bf B246} (1990) 377
%%CITATION=PHLTA,B246,377;%%
\semi
J.D. Lykken, {\it Weak Scale Superstrings,} Phys. Rev. {\bf D54} (1996)
3693, {\tt hep-th/9603133}
%%CITATION=PHRVA,D54,3693;%%
\semi
K.R. Dienes, E. Dudas and T. Gherghetta, {\it Extra Spacetime Dimensions and
Unification,} Phys. Lett. {\bf B436} (1998) 55 {\tt hep-th/9803466}
%%CITATION=PHLTA,B436,55;%%
\semi
K.R. Dienes, E. Dudas and T. Gherghetta, {\it Grand Unification at Intermediate
Mass Scales through Extra Dimensions,} Nucl. Phys. {\bf B537} (1999) 47,
{\tt hep-th/9806292}
%%CITATION=NUPHA,B537,47;%%
\semi
Z. Kakushadze, {\it TeV Scale Supersymmetric Standard Model and Brane
World,} {\tt hep-th/9812163}.}
%%CITATION=HEP-TH 9812163;%%
\lref\rNeg{For a recent review see for example: 
J.W. Negele, {\it Instantons, the QCD Vacuum and Hadronic Physics,}
{\tt hep-lat/9810153}.}
%%CITATION=HEP-LAT 9810153;%%
\lref\rLoop{S-J. Rey and J. Yee, {\it Macroscopic Strings as heavy quarks
in Large N Gauge Theory,} {\tt hep-th/9803001}
%%CITATION=HEP-TH 9803001;%%
\semi
J. Maldacena, {\it Wilson Loops in Large N Field Theories,}
Phys. Rev. Lett. {\bf 80} (1998) 4859, {\tt hep-th/9803002}
%%CITATION=PRLTA,80,4859;%%
\semi
E. Witten {\it Anti-de Sitter Space, Thermal Phase Transition and Confinement
in Gauge Theories,} Avd. Theor. Math. Phys. {\bf 2} (1998) 505
{\tt hep-th/9803131}.}
%%CITATION=HEP-TH 9803131;%%
\lref\rIgK{I.R. Klebanov and A.A. Tseytlin, {\it D Branes and Dual Gauge 
Theories in Type-0 Strings,} {\tt hep-th/9811035}
%%CITATION=HEP-TH 9811035;%%
\semi
J.A. Minahan, {\it Glueball Mass Spectra and other issues for Supergravity
Duals of QCD Models,} {\tt hep-th/9811156}
%%CITATION=HEP-TH 9811156;%%
\semi
J.A. Minahan, {\it Asymptotic Freedom and Confinement from Type 0 String
Theory} {\tt hep-th/9902074}
%%CITATION=HEP-TH 9902074;%%
\semi
I.R. Klebanov and A.A. Tseytlin, {\it Asymptotic Freedom and Infrared
Behaviour in the Type-0 String Approach to Gauge Theory,}
{\tt hep-th/9812089.}}
%%CITATION=HEP-TH 9812089;%%
\lref\rWK{For nice discussions of this see: E. Witten, 
{\it New Perspectives in the Quest for Unification,} 
{\tt hep-ph/9812208}
%%CITATION=HEP-PH 9812208;%%
\semi
I.R. Klebanov, {\it From Threebranes to Large N Gauge
Theory,} {\tt hep-th/9901018.}}
%%CITATION=HEP-TH 9901018;%%
\lref\rSon{A. Brandhuber, N. Itzhaki, J. Sonnenschein and S. Yankielowicz,
{\it Wilson Loops in the Large N Limit at Finite Temperature,}
Phys. Lett. {\bf B434} (1998) 36, {\tt hep-th/9803137.}}
%%CITATION=PHLTA,B434,36;%%
\lref\rMal{J. Maldacena, {\it The Large N Limit of Superconformal
Field Theories and Supergravity,} Adv.Theor.Math.Phys. {\bf 2} (1998) 231,
{\tt hep-th/9711200}
%%CITATION=HEP-TH 9711200;%%
\semi
E. Witten, {\it Anti-de Sitter Space and Holography,} Adv.Theor.Math.Phys.
{\bf 2} (1998) 253, {\tt hep-th/9802150}
%%CITATION=HEP-TH 9802150;%%
\semi
S.S. Gubser, I.R. Klebanov and A.M. Polyakov, {\it Gauge Theory Correlators
from Noncritical String Theory,} Phys.Lett. {\bf B428} (1998) 105, 
{\tt hep-th/9802109}.}
%%CITATION=PHLTA,B428,105;%%
\lref\rTh{G. 't Hooft, {\it A Planar Diagram Theory for Strong Interactions,}
Nucl. Phys. {\bf B72} (1974) 461.}
%%CITATION=NUPHA,B72,461;%%
\lref\rWS{N. Seiberg and E. Witten, {\it Electric-Magnetic Duality, Monopole
Condensation and Confinement in N=2 Super Yang-Mills Theory,} 
Nucl. Phys. {\bf B426} (1994) 19, {\tt hep-th/9407087.}}
%%CITATION=NUPHA,B426,19;%%
\lref\rPol{A. Polyakov, {\it The Wall of the Cave,} {\tt hep-th/9809057.}}
%%CITATION=HEP-TH 9809057;%%
\lref\rES{E. Witten and L. Susskind, {\it The Holographic Bound in Anti-de 
Sitter Space,} {\tt hep-th/9805114.}}
%%CITATION=HEP-TH 9805114;%%
\lref\rGuby{A. Kehagias and K. Sfetsos, {\it On Running Couplings in Gauge
Theories from IIB supergravity,} {\tt hep-th/9902125}
%%CITATION=HEP-TH 9902125;%%
\semi
S.S. Gubser, {\it Dilaton-driven Confinement,} {\tt hep-th/9902155.}}
%%CITATION=HEP-TH 9902155;%%
\lref\rUs{R. de Mello Koch, A. Paulin-Campbell and J.P. Rodrigues, work
in progress.}
\lref\rbuch{I.L. Buchbinder and S.M. Kuzenko, {\it Comments on the Background
Field Method in Harmonic Superspace: Nonholomorphic Corrections in N=4
super Yang-Mills,} Mod. Phys. Lett. {\bf A13} (1998) 1623, 
{\tt hep-th/9804168}
%%CITATION=MPLAE,A13,1623;%%
\semi
E.L. Buchbinder, I.L. Buchbinder and S.M. Kuzenko, {\it Nonholomorphic
Effective Potential in N=4 SU(n)
SYM,} Phys. Lett. {\bf B446} (1999) 216, {\tt hep-th/9810239}.}
%%CITATION=PHLTA,B446,216;%%
\lref\rSpa{A. Fayyazuddin and M. Spalinski, {\it Large N Superconformal Gauge
Theories and Supergravity Orientifolds,} Nucl. Phys. {\bf B535} (1998) 219,
{\tt hep-th/9805096.}}
%%CITATION=NUPHA,B535,219;%%
\lref\rsen{A.Sen, {\it F Theory and Orientifolds,} Nucl. Phys. {\bf B475}
(1996) 562, {\tt hep-th/9605150.}}
%%CITATION=NUPHA,B475,562;%%

%%%%%%%%%%%%%%%%%%%%%%%%%%%%%%%%%%%%%%%%%%%%%%%%%%%%%%%%%%%%%%%%%%%%%%
%%%%%%%%%%%%%%%%%%          title page       
%%%%%%%%%%%%%%%%%%%%%%%%%%%%%%%%%%%%%%%%%%%%%%%%%%%%%%%%%%%%%%%%%%%%%%

\Title{ \vbox {\baselineskip 12pt\hbox{CNLS-99-02}
\hbox{BROWN-HET-1174}  \hbox{March 1999}  }}
{\vbox {\centerline{Non-holomorphic Corrections from Threebranes}
        \centerline{in F Theory}
}}

\smallskip
\centerline{Robert de Mello Koch$^{1}$, 
Alastair Paulin-Campbell$^{2}$ and Jo\~ao P. Rodrigues$^{2}$}
\smallskip
\centerline{\it Department of Physics,$^{1}$}
\centerline{\it Brown University}
\centerline{\it Providence RI, 02912, USA}
\centerline{\tt robert@het.brown.edu}\bigskip

\smallskip

\centerline{\it Department of Physics and Center for Nonlinear Studies,$^{2}$}
\centerline{\it University of the Witwatersrand,}
\centerline{\it Wits, 2050, South Africa}
\centerline{\tt paulin,joao@physnet.phys.wits.ac.za}\bigskip

\noindent
We construct solutions of type IIB supergravity corresponding to $7$ branes, an $O7$ plane
and 3 branes. 
By considering a probe moving in this background,
with constant coupling and an AdS$_{5}$ component in its geometry, we are
able to reproduce the exact low energy effective action for ${\cal N}=2$ super Yang-Mills
theory with
gauge group $SU(2)$ and $N_{f}=4$ massless flavors. After turning on a mass
for the flavors we find corrections to the AdS$_{5}$ geometry. In addition, 
the coupling shows a power law dependence on the energy scale of the theory.
The origin of the power law behaviour of the coupling is traced back to
instanton corrections. Instanton corrections to the four derivative terms 
in the low energy effective action are correctly obtained from a probe 
analysis. We study how these instanton corrections are reflected in the background geometry
by calculating the quark-antiquark potential.
Finally we consider a solution corresponding to an asymptotically free
field theory. Again, the leading form of the four derivative terms in the
low energy effective action are in complete agreement with field theory 
expectations.

%%%%%%%%%%%%%%%%%%%%%%%%%%%%%%%%%%%%%%%%%%%%%%%%%%%%%%%%%%%%%%%%%%%%%%

\Date{}

%%%%%%%%%%%%%%%%%%%%%%%%%%%%%%%%%%%%%%%%%%%%%%%%%%%%%%%%%%%%%%%%%%%%%%
%%%%%%%%%%%%                text begins                        
%%%%%%%%%%%
%%%%%%%%%%%%%%%%%%%%%%%%%%%%%%%%%%%%%%%%%%%%%%%%%%%%%%%%%%%%%%%%%%%%%%

\newsec{Introduction}

Recent progress in string theory has lead to a deep and powerful connection
between Yang-Mills theory and string theory, something which was expected
more than twenty years ago\rTh. In particular, according to
Maldacena\rMal\ the large $N$ and large 't Hooft coupling dynamics of 
Yang-Mills theory is captured by supergravity. This relationship
can be motivated by studying a brane in full string theory. One then considers
a low energy limit which decouples the field theory from gravity, and at
the same time one considers the near horizon limit of the corresponding
supergravity solution.

In this paper, we study the low energy limit of ${\cal N}=2$ super
Yang-Mills theory, by realizing it as the world volume theory on a Dirichlet
three brane moving near seven branes, that is, threebranes in F theory.
In this way we provide an alternative field theory interpretation which is 
obtained by probing with this brane the near horizon supergravity solutions. 
We begin by identifying the exact effective complex coupling\rWS\ of the 
low energy field theory with the complex coupling of type IIB 
supergravity. The remaining supergravity equations then determine a unique
metric. This supergravity solution corresponds to a sevenbrane background.
We then consider introducing a large number $N$ of coincident threebranes 
into the problem. The supergravity solution for the threebranes plus sevenbranes 
has the
same complex coupling as the pure sevenbrane background\rOfer. The 
presence of the threebranes does however
deform the geometry and switch on a flux for the self-dual five form. The
flux and deformed metric are determined by solving the Laplace equation on
the background generated by the sevenbranes\rOfer. In the large $N$ limit 
('t Hooft limit) both curvature and string loop corrections to the background
are suppressed.
The field theory of 
interest is then realized as a Born Infeld action describing the worldvolume dynamics
of a threebrane probe moving
in this geometry. In the supergravity description, we are studying the two-body interaction
between the source and probe threebranes. We will thus compare the worldvolume
theory of the probe to the low energy effective action for the field theory 
with gauge group $SU(2)$ as suggested by the work of \rHyun. 

The first example we consider is the theory with $N_{f}=4$ massless flavors.
This theory is a finite conformal field theory and as expected the geometry
contains an AdS$_{5}$ factor. By requiring that the effective action in field 
theory has an exact $SL(2,Z)$ symmetry, we are able to fix the form of the six 
derivative terms\rocek. After performing a field redefinition on the field
theory side \rocek,\rJev,\rHo, the form of the field theory 
effective action and probe
action agree up to and including six derivative contributions. This 
computation is a simple
extension of the result reported in \rocek\ for the ${\cal N}=4$ case. The 
new feature is the identification of the explicit $N$ dependence of the 
background coupling which is needed to exhibit the expected overall $N$
dependence of the probe action. 

The next example we study is the theory with $N_{f}=4$ massive flavors. The 
presence of the flavor masses explicitly breaks the conformal invariance. 
Indeed, the coupling of this theory picks up a dependence on the energy scale 
of the theory due to instanton corrections. Solutions of this type are
particularly interesting from the point of view of the gravity/field theory
connection. The radial direction of the AdS$_{5}$ space plays the role of an
energy scale in the conformal field theory case\rES. However, all
beta functions in the conformal field theory vanish and the evolution under
the renormalization group is trivial. Examples of quantum field theories that
are not conformal have a much richer renormalization group evolution and may
provide valuable insights into the role of the "fifth dimension"\rPol. In
this case, we are able to determine the first corrections to the AdS$_{5}$
geometry. This approximate background leads to corrections to the four
derivative terms in the probe action that are consistent with the form of 
instanton corrections to the four derivative terms in the field theory.
Our results strongly suggest that the probe worldvolume action reproduces the 
exact Wilsonian effective action of the field theory. 
We also compute the quark-antiquark potential in this geometry.

Our third and final example considers the pure gauge theory. We are able
to find an approximate solution, valid when the separation between the probe
and source is large. This geometry reproduces the correct {\it semiclassical}
structure of the four derivative terms. It would thus seem that the background
we have found is capable of describing {\it perturbative} field theory on
the probe world volume. By explicitly computing the magnitude of the
square of the Ricci tensor, we are able to explain why this is indeed 
the case.  

When this work was near completion we received \rGuby, \rNojiri. In these papers
solutions of IIB supergravity corresponding to non-conformal field theories
were constructed. These solutions preserve the $SO(6)$ invariance of the
AdS$_{5}\times$S$_{5}$ solution, and thus are similar but not identical to
the solution we obtain in the $N_{f}=4$ massless flavors case.

\newsec{${\cal N}=2$ Super Yang-Mills Theory with gauge group $SU(2)$ and 
$N_{f}=4$ Massless Multiplets}

In this section we begin by collecting the known field theory results on
the low energy effective action of ${\cal N}=2$ super Yang-Mills theory
with gauge group $SU(2)$ and $N_{f}=4$ massless matter multiplets. These 
are then compared to the result obtained from studying a probe threebrane
in supergravity.

\subsec{Field Theory Results}

The perturbative beta function for ${\cal N}=2$ super Yang-Mills theory with 
gauge group $SU(N_{c})$ and $N_{f}$ flavor hypermultiplets is proportional to
$(2N_{c}-N_{f})$. Thus, for $N_{c}=2$ and $N_{f}=4,$ the perturbative
beta function vanishes. If in addition all of the flavors of matter are
massless, we obtain a finite conformally invariant theory\rSW. The exact
effective coupling of the theory has the form 

$$ \tau=\tau_{1}+i\tau_{2}=\tau_{cl}
+{i\over\pi}\sum_{n=0,2,4,...}c_{n}e^{in\pi\tau_{cl}}
=\tau_{cl}+{i\over\pi}\sum_{n=0,2,4,...}c_{n}q^{n},$$

\noindent
where $\tau_{cl}$ is the 
classical coupling of the theory. The coefficient $c_{0}$
is a one loop perturbative correction, which in the Pauli-Villars scheme, has
the value $c_{0}=4\log(2)$\rdorey. The coefficients $c_{n}$ with $n>0$ and 
even come from nonperturbative (instanton) effects. The two instanton 
coefficient has been computed and has the value $c_{2}=-7/(2^{6}3^{5})$. The
leading contribution to the low energy effective action comprises all terms
with the equivalent of two derivatives or four fermions\rHen\ and is 
determined in terms of the effective coupling. The next-to-leading contribution 
to the low energy effective action contains all terms with the equivalent of
four derivatives or eight fermions\rHen. Using the scale invariance and
$U(1)_{\cal R}$ symmetry of the model, Dine and Seiberg argued that the four 
derivative term is one loop exact\rDS. In \rKD\ the vanishing of instanton
corrections to the four derivative terms was explicitly verified and a 
rigorous proof of this non-renormalization theorem has recently been given 
in \rOvrut. The one loop contribution to the four derivative terms has been
considered in \rOLoop. The result for the low energy effective action, up to 
and including four derivative terms, in ${\cal N}=2$ superspace, is given by 

\eqn\LEEA
{8\pi\big(S_{eff}^{(2)}+S_{eff}^{(4)}\big)=
{\cal I}m\int d^{4}xd^{4}\theta\big({1\over 2}A^{2}\big)+
{3\over 128\pi^{2}}\int d^{4}xd^{4}\theta 
d^{4}\bar{\theta}\log A\log\bar{A},}

\noindent
where $A$ is an ${\cal N}=2$ Abelian chiral superfield. The number of terms
that contribute to the low energy effective action at each order, for six
derivative terms or higher, increases rapidly and a direct approach to these
terms is not feasible. An elegant approach to study these terms has
been developed in \rocek\ for ${\cal N}=4$ super Yang-Mills\rbuch, based on the 
conjectured $SL(2,Z)$ duality of the theory. This duality was used to fix the 
form of the effective action
up to six derivatives. The theory that we are studying is also believed to have
an exact $SL(2,Z)$ duality\rSW, and under this assumption the analysis of 
\rocek applies. 

\noindent
The unique $SL(2,Z)$ invariant form for the six
derivative terms is

\eqn\SixDTer
{\eqalign{8\pi S_{eff}^{6}=
\Big({3\over 128 \pi^{2}}\Big)^{2}
\lambda^{(6)}\int &d^{4}x d^{4}\theta d^{4}\bar{\theta}
\Big({1\over\sqrt{\tau\bar{\tau}}}
{\bar{D}^{\dot{\alpha}}{}_{a}
\bar{D}_{\dot{\alpha} b}\log (\bar{A})\over A}
{D_{\alpha}{}^{a}D^{\alpha b}\log (A)\over\bar{A} }\Big)\cr
+{i\over 2}\Big({3\over 128\pi^{2}}\Big)^{2}
\int &d^{4}x d^{4}\theta d^{4}\bar{\theta}
\Big(\log(\bar{A})
{\bar{D}_{\dot{\alpha}1}\bar{D}^{\dot{\alpha}}{}_{1}\bar{D}_{\dot{\beta} 2}
\bar{D}^{\dot{\beta}}{}_{2}\log\bar{A}\over \tau A^{2}}\cr
&\qquad-
\log (A){D_{\alpha}{}^{1}D^{\alpha 1}D_{\beta}{}^{2}
D^{\beta 2}\log (A)\over \bar{\tau}\bar{A}^{2}}\Big),}}

\noindent
Under duality, the second term above mixes with the two and four derivative
terms and consequently its coefficient is fixed. The requirement of self
duality does not fix $\lambda^{(6)}$, since duality maps this term into itself
at lowest order. These are the field theory results that we wish to compare to 
gravity.

On the gravity side, we will consider a probe moving in a background to
be specified below. The probe worldvolume dynamics is captured by a Born-Infeld
action. The Born-Infeld action itself is self-dual, but the duality does not
act on the separation of the branes. This separation is parametrized by the
Higgs fields which belong to the same supermultiplet as the gauge fields.
This implies, as pointed out in \rocek,  that the Higgs fields that realize
${\cal N}=2$ supersymmetry linearly must be related by a nonlinear gauge field
dependent redefinition to the separation. It is interesting to note that a
similar field redefinition is needed to map the linear realization of
conformal symmetry in super Yang-Mills theory into the isometry of the 
Anti de-Sitter spacetime of the supergravity description\rJev. We refer
the reader to \rocek\ for the detailed form of the field redefinitions.
The result after performing the field redefinitions, in terms of component 
fields, reads

\eqn\Reslt
{\eqalign{S_{eff}=&\int d^{4}x
\Big(-{1\over 4g^{2}}\partial_{m}\bar{\varphi}
\partial^{m}\varphi-{1\over 8g^{2}}(F_{\alpha\beta}F^{\alpha\beta}
+\bar{F}_{\dot{\alpha}\dot{\beta}}\bar{F}^{\dot{\alpha}\dot{\beta}})
+\Big({3\over 128\pi^{2}}\Big)^{2}\times\cr
&\times {1\over 32\pi}{F_{\alpha\beta}F^{\alpha\beta}
\bar{F}_{\dot{\alpha}\dot{\beta}}\bar{F}^{\dot{\alpha}\dot{\beta}}
+(\partial_{m}\varphi\partial^{m}\varphi)(\partial_{n}\bar{\varphi}
\partial^{n}\bar{\varphi})-F^{\beta\alpha}\partial_{m}\varphi
\sigma^{m}{}_{\alpha\dot{\beta}}\bar{F}^{\dot{\beta}\dot{\alpha}}
\partial_{n}\bar{\varphi}\sigma^{n}{}_{\beta\dot{\alpha}}\over
\varphi^{2}\bar{\varphi}^{2}}\cr
&-{g^{2}\over 256\pi^{2}}
\Big({3\over 128\pi^{2}}\Big)^{2}
{F_{\alpha\beta}F^{\alpha\beta}
\bar{F}_{\dot{\alpha}\dot{\beta}}\bar{F}^{\dot{\alpha}\dot{\beta}}
(F^{\rho\tau}F_{\rho\tau}+\bar{F}^{\dot{\rho}\dot{\tau}}
\bar{F}_{\dot{\rho}\dot{\tau}})
\over\varphi^{4}\bar{\varphi}^{4}}
\Big),}}

\noindent
where we have set $\tau=i{4\pi\over g^{2}}.$ The six derivative terms for 
the scalars are not displayed since they depend on the arbitrary constant
$\lambda^{(6)}$. The value of $\lambda^{(6)}$ as well as the structure of
the effective action given above can be checked by explicitly computing
instanton corrections to the six derivative terms. We hope to return to this
in the near future\rUs. Notice that all acceleration terms were eliminated by 
the field redefinition, something first noted in\rHo.

\subsec{Supergravity Results}

The supergravity background relevant for the study of ${\cal N}=2$ 
supersymmetric field theory is generated by sevenbranes and a large number of 
threebranes\rOfer, i.e. threebranes in F theory\foot{Supergravity backgrounds 
corresponding to ${\cal N}=1$ field theories have been considered in \rAhn.}. 
To construct this 
background it is convenient to start with a solution for the sevenbranes 
by themselves\rOfer. The sevenbrane solution is described in terms of 
non-zero metric, dilaton and axion fields. As usual, the metric and axion are
combined into a single complex coupling 
$\tau=\chi+ie^{-\phi}=\tau_{1}+i\tau_{2}$. The coupling
$\tau$ is identified with the modular parameter of the elliptic fiber of the
F theory compactification. The $(8,9)$ plane is taken to be orthogonal to the
sevenbranes. In terms of the complex coordinate
$z=x^{8}+ix^{9}$ we make the following ansatz for the metric

\eqn\frst
{ds^{2}=e^{\varphi(z,\bar{z})}dzd\bar{z}+(dx^{7})^{2}+...
+(dx^{1})^{2}-(dx^{0})^{2}.}

\noindent
The parameter $z$ is to be identified with the Higgs field appearing
in the low energy effective action of the ${\cal N}=2$ field theory.
With this ansatz, the type IIB supergravity equations of 
motion reduce to\rGSVY\

\eqn\scnd
{\eqalign{\partial\bar{\partial}\tau &=
{2\partial\tau\bar{\partial}\bar{\tau}\over\bar{\tau}-\tau}\cr
\partial\bar{\partial}\varphi &=
{\partial\tau\bar{\partial}\bar{\tau}\over(\bar{\tau}-\tau)^{2}}.}}

\noindent
The complex coupling $\tau$ is identified with the low energy effective 
coupling of the ${\cal N}=2$ field theory. Supersymmetry constrains the
effective coupling of the field theory to be a function of $z$, so that
the first equation in \scnd\ is automatically satisfied. The general 
solution to the second equation in \scnd\ is 

\eqn\thrd
{\varphi (z,\bar{z})=\log(\tau_{2}) +F(z)+\bar{F}(\bar{z}).}

\noindent
The functions $F(z)$ and $\bar{F}(\bar{z})$ should be chosen in order that
\frst\ yields a sensible metric. For the case that we are considering,
the explicit form for the metric transverse to the sevenbranes is

\eqn\htgdgf
{ds^{2}=e^{\varphi (z,\bar{z})}dzd\bar{z}=\tau_{2}|da|^{2},}

\noindent
where $a$ is the quantity that appears in the Seiberg-Witten solution\rOfer.
This specifies the solution for the sevenbranes by themselves.

Next following \rOfer, we introduce threebranes into the problem\foot{See 
also \rKeh\ where this solution was independently discovered.}. The presence
of the threebranes modifies the metric and switches on a non-zero flux for
the self dual RR five-form field strength. The world volume coordinates of 
the threebranes are  $x^{0},x^{1},x^{2},x^{3}$. One obtains a valid 
solution\rOfer\ by making the following ansatz for the metric

\eqn\frth
{ds^2 = f^{-1/2}dx_{\parallel}^2 + f^{1/2}{g}_{ij}dx^{i}dx^{j}}

\noindent
and the following ansatz for the self-dual 5-form field strength

\eqn\ffth
{F_{0123i}  =  -{{1}\over{4}}\partial_{i}f^{-1}~. }

\noindent
The complex field $\tau$ is unchanged by the introduction of the threebranes.
Inserting the above ansatz into the IIB supergravity equations of motion, 
one finds that $f$ satisfies the following equation of motion

\eqn\sxh
{ {1\over\sqrt{g}} 
\partial_{i}(\sqrt{ g}{ g}^{ij}\partial_j f)=
- (2 \pi)^4 N { \delta^6(x-x^0) \over \sqrt{\ g}. }}

\noindent
This last equation corresponds to the case in which all of the three branes
are located at the same point. In the limit that $N\to\infty$ the curvature
becomes small almost everywhere and the supergravity solution can be used 
to reliably compute quantities in the field theory limit as explained 
in\rOfer. 

In the case of $N_f=4$ massless hypermultiplets, \sxh\ is explicitly
given by

\eqn\explicit
{\big[\tau_{2}\partial_{y}^{2}+4\partial_{a}\partial_{\bar{a}}\big]
f=-(2\pi )^{4}N\delta^{(4)}(y)\delta^{(2)}(a).}

\noindent
The solution is given by
\foot{It is well known\rsen\ the case of constant $\tau$ corresponds to a IIB
orientifold background. Threebranes in a IIB orientifold background were
first considered in \rSpa.} 

$$ f={4N\pi\over\big[y^{2}+\tau_{2}|a|^{2}\big]^{2}}. $$

To reproduce the low energy effective action of the field theory, we now
consider the dynamics of a threebrane probe moving in this geometry. 
It is well known that the probe has a low energy effective action which matches
that of the corresponding low energy field theories\rDoug. Here we are
interested in checking the form predicted by the probe for the higher order
corrections. The leading low energy 
effective action plus corrections for the bosons in the background
described above, is obtained by expanding the self-dual action \rDaction:

\eqn\svnth
{S={T_{3}\over 2}
\int d^{4}x\Big[\sqrt{det(G_{mn}+e^{-{1\over 2}\phi}F_{mn})}+\chi
F\wedge F\Big].}

\noindent
$T_3$ has no dependence on the string coupling constant. We obtain for the scalar terms obtained from the expansion of \svnth\ after setting $y=0$:

\eqn\svnthexp
{\eqalign{S&\sim {1\over 2}\int d^{4}x
\Big(\tau_{2}\partial_{m}a\partial^{m}\bar{a}
-{f\over 2}\tau_{2}^{2}(\partial_{m}a\partial^{m}a)
(\partial_{n}\bar{a}\partial^{n}\bar{a})\cr
&+{f^{2}\over 2}\tau_{2}^{3}(\partial_{p}a\partial^{p}\bar{a})
(\partial_{m}a\partial^{m}a)
(\partial_{n}\bar{a}\partial^{n}\bar{a})+ ...\Big)\cr
&={1\over 2}\int d^{4}x \Big(
\tau_{2}\partial_{m}a\partial^{m}\bar{a}
-{2N\pi\over (a\bar{a})^{2}}(\partial^{m}a\partial_{m}a) 
(\partial^{n}\bar{a}\partial_{n}\bar{a})\cr
&\qquad +{8\pi^{2}N^{2}\over \tau_{2}(a\bar{a})^{4}}
(\partial_{p}a\partial^{p}\bar{a})
(\partial^{m}a\partial_{m}a) 
(\partial^{n}\bar{a}\partial_{n}\bar{a})+ ...\Big).}}

\noindent
Notice that each term in this action comes multiplied by a different power of 
$N$. As things stand, the $2n$ derivative term will come with a coefficient
of $\tau_{2}^{n}f^{n-1}\sim N^{n-1}$. The full effective action for the probe
interacting with $N$ coincident source threebranes, should come with an overall
factor of $N$\rTseyt. This is achieved by noting that the coupling of the 
background, $\tau_{2}$ should be identified with

\eqn\Rel
{\tau_{2}\equiv {N\tau_{2,SW}\over\lambda} = {1 \over g_s},}

\noindent
where $\tau_{SW}$ is the Seiberg-Witten effective coupling for the field 
theory of interest. 
With this identification, the string coupling is $O({1\over N})$ and  
explicitly goes to zero as $N\to\infty$. Notice also that 
$Ng_{s}=Ng_{YM}^{2}\sim\lambda$ so that the large 't Hooft coupling limit
corresponds to large $\lambda$. After making the $N$ dependence of $\tau_{2}$
explicit, we find that the probe action is indeed proportional to $N$.
We will not always show this dependence explicitly in what follows.
Following \rocek, we find the Taylor expansion of \svnth\ exactly 
matches the super Yang-Mills effective action \Reslt\  
after identifying

$$  a\bar{a}={1\over T_{3}\lambda}\varphi\bar{\varphi},\qquad
    (F_{s,\alpha\beta}F_{s}{}^{\alpha\beta})=
    {1\over 4T_{3}\lambda}
    (F_{f,\alpha\beta}F_{f}{}^{\alpha\beta})  $$

\noindent
where $F_{s,\alpha\beta}$ is the field strength appearing on the probe
worldvolume and $F_{f,\alpha\beta}$ is the field strength of the field
theory. Note that $\tau_{2}$ appearing in \svnthexp\ is the classical 
coupling plus all instanton corrections. The fact that the four derivative 
terms are independent of $\tau_{2}$ shows that the supergravity result
explicitly reproduces the nonrenormalization theorem for these terms\rBob. 

\newsec{${\cal N}=2$ Super Yang-Mills Theory with gauge group $SU(2)$ and 
$N_{f}=4$ Massive Multiplets}

In this section we consider the supergravity background corresponding to 
the case where all flavor multiplets of the field theory on the probe world
volume have a mass. In this case, both the
effective coupling and the four derivative terms get contributions from 
instantons. We are able to show that the supergravity solution is capable
of producing what is expected for the one instanton correction. We are not
however able to fix the coefficient of this correction. The dilaton
of the supergravity solution is no longer a constant and there are corrections
to the AdS$_{5}$ geometry reflecting the fact that the field theory is no 
longer conformally invariant. We compute the quark-antiquark potential and
show that its form is remarkably similar to that for a quark-antiquark pair 
in the ${\cal N}=4$ theory at finite temperature.

\subsec{Field Theory Results}

The masses of the quark flavors breaks the conformal invariance that is
present in massless theory. In this case, the effective coupling does 
pick up a dependence on the energy scale as a result of instanton corrections.
At high enough energies we expect these corrections can be neglected and
the theory flows to the conformal field theory corresponding to the case of
massless flavors. Indeed, the perturbative beta function still vanishes and
the coupling goes to a constant at high energies. We will focus attention
on the two and four derivative terms appearing in the low energy effective
action. These terms are completely specified by a holomorphic prepotential
${\cal F}$ and a real function ${\cal H}$

$${S_{eff}={1\over 2i}\int d^{4}x\Big(\int d^{4}\theta{\cal F}(A)
-\int d^{4}\bar{\theta}\bar{{\cal F}}(\bar{A})\Big)+\int d^{4}x
\int d^{4}\theta d^{4}\bar{\theta}{\cal H}(A,\bar{A}).}$$

\noindent
In what follows, we will only account for the one instanton corrections to 
both the prepotential and the four derivative terms. The prepotential does
not receive any loop corrections for $N_f =4$. The one instanton 
correction to the prepotential was computed in \rKh. The one instanton 
corrected prepotential is

$$ {\cal F}={1\over 2}\tau_{cl}A^{2}-{i\tau_{cl}\over 2\pi}
   {q\over A^{2}}m_{1}m_{2}m_{3}m_{4}. $$

\noindent
This corresponds to a low energy effective coupling

\eqn\LEC
{\tau=\tau_{cl}-{3iq\tau_{cl}\over\pi \varphi^{4}}m_{1}m_{2}m_{3}m_{4}}

\noindent
The one loop correction to the real function ${\cal H}$ is\rOLoop\

$$ {\cal H}= {3\over 256\pi^{2}}\log^{2}\Big(
   {A\bar{A}\over\langle A\rangle \langle \bar{A}\rangle}\Big), $$

\noindent 
and the one instanton correction is given by

$$ {\cal H}(\varphi,\bar{\varphi})={-qm_{1}m_{2}m_{3}m_{4}\over
   8\pi^{2}\varphi^{4}} \log\bar{\varphi}.$$

\noindent
The one anti-instanton contribution is given by the complex conjugate
of the one instanton correction. The pure scalar two and four derivative 
terms appearing in the low energy effective action, after performing the
field redefinition needed to compare to the brane result, are easily
obtained by using the formulas quoted in \rHo,\rBob. The results are

\eqn\LowEn
{S=\int d^{4}x\Big( K_{\varphi\bar{\varphi}}
\partial_{\mu}\varphi\partial^{\mu}\bar{\varphi}+
\tilde{\cal H}_{\varphi\varphi\bar{\varphi}\bar{\varphi}}
(\partial^{m}\varphi)
(\partial_{m}\varphi)(\partial^{n}\bar{\varphi})
(\partial_{n}\bar{\varphi})\Big),}

\noindent
where

$$  K_{\varphi\bar{\varphi}}\equiv Im\Big(
    {\partial^{2}{\cal F}\over\partial\varphi^{2}}\Big) 
    =\tau_{2}={4\pi^{2}\over g_{cl}^{2}}-
{6\pi\over g_{cl}^{2}}m_{1}m_{2}m_{3}m_{4}\Big[
{q\over a^4}+{\bar{q}\over \bar{a}^4}\Big]$$

\noindent
and

$$ {\eqalign{\tilde{\cal H}_{\varphi\varphi\bar{\varphi}\bar{\varphi}}&=
  16\Big({\partial^{4}{\cal H}\over \partial\varphi\partial\varphi
  \partial\bar{\varphi}\partial\bar{\varphi}}-
  {\partial^{3}{\cal H}\over \partial\varphi\partial\varphi
  \partial\bar{\varphi}}
  (K_{\bar{\varphi}\varphi})^{-1}{\partial 
  K_{\varphi\bar{\varphi}}\over\partial\bar{\varphi}}-
  {\partial K_{\varphi\bar{\varphi}}\over \partial\varphi}
  (K_{\bar{\varphi}\varphi})^{-1}
  {\partial^{3}{\cal H}\over \partial\varphi
  \partial\bar{\varphi}\partial\bar{\varphi}}\cr
  &+2{\partial K_{\varphi\bar{\varphi}}\over\partial\varphi}
  (K_{\bar{\varphi}\varphi})^{-1}
  {\partial^{2}{\cal H}\over \partial\varphi\partial\bar{\varphi}}
  (K_{\bar{\varphi}\varphi})^{-1}{\partial K_{\varphi\bar{\varphi}}
  \over\partial\bar{\varphi}}\Big)\cr
  &={3\over 8\pi^{2}}
  {1\over\varphi^{2}\bar{\varphi}^{2}}
  +{40 m_{1}m_{2}m_{3}m_{4}\over\pi^{2}}
  \Big[{q\over\bar{\varphi}^{2}\varphi^{6}}
  +{\bar{q}\over\varphi^{2}\bar{\varphi}^{6}}\Big].}} $$

\noindent
Notice that for large $\varphi$ the fall off of the four derivative terms
is like $|\varphi|^{-4}$. This has an interesting supergravity interpretation.

\subsec{Supergravity Results}

The first step in the supergravity analysis entails solving \sxh\ for the
background geometry, with the complex coupling $\tau$ given in \LEC. The
coupling $\tau$ is only valid for large $|\varphi|$. For small $|\varphi|$
higher instanton corrections can not be neglected. For this reason, we will
construct a solution to \sxh\ which is valid for large $|\varphi|$. Towards
this end, split $\tau_{2}$ into two pieces as follows

$$ \tau_{2}=V_{1}-V_{2},\quad V_{1}={4\pi^{2}\over g_{cl}^{2}}
   \equiv\tau_{2cl},\quad V_{2}=
   {3\tau_{2cl}\over 2\pi}m_{1}m_{2}m_{3}m_{4}\Big[
   {q\over a^4}+{\bar{q}\over \bar{a}^4}\Big]. $$

\noindent
We can now solve \sxh\ perturbatively by writing $f=f_{0}+f_{1}+...$ where

\eqn\FrtPbn
{\Big[V_{1}{\partial^{2}\over\partial y^{2}}+{\partial^{2}\over\partial a\partial\bar{a}}\Big]
f_{0}=-N(2\pi )^{4}\delta^{(4)}(y)\delta^{(2)}(a),}

\eqn\SndPbn
{\Big[V_{1}{\partial^{2}\over\partial y^{2}}+{\partial^{2}\over\partial a\partial\bar{a}}\Big]
f_{n}=V_{2}{\partial^{2}\over\partial y^{2}}f_{n-1}.}

\noindent
To find the leading corrections to the four derivative terms, it is
sufficient to focus attention on $f_{0}$ and $f_{1}$. The solution for
$f_{0}$ is 

$$ f_{0}={4N\pi\over\big[y^{2}+\tau_{2cl}|a|^{2}\big]^{2}}. $$

\noindent
The function $f_{1}$ satisfies

$$  \Big[\tau_{2cl}{\partial^{2}\over\partial y^{2}}+
    {\partial^{2}\over\partial a\partial \bar{a}}\Big]f_{1}=
    {3m_{1}m_{2}m_{3}m_{4}\tau_{2cl}\over 2\pi}
    \Big({q\over a^{4}}+{\bar{q}\over\bar{a}^{4}}\Big)
    \Big(-{16N\pi\over \big[y^{2}+\tau_{2cl}a\bar{a}\big]^{3}}
    +{96N\pi y^{2}\over \big[y^{2}+\tau_{2cl}a\bar{a}\big]^{4}}
    \Big). $$

\noindent
We will look for solutions to this equation that preserve rotational symmetry 
in the $y_{i}$ variables. To do this it is useful to move into radial 
coordinates. Denoting the angular variable in the $a,\bar{a}$ plane by 
$\theta$ and the radial coordinate in the the $a,\bar{a}$ plane by $r$ and 
in the $y_{i}$ plane by $\rho$, we find\foot{The factors $q$ and $\bar{q}$
appearing in $V_{2}$ are pure phases and can be absorbed into a convenient
choice for $\theta=0$.}

$$ {\eqalign{\Big[\tau_{2cl}&
   {\partial^{2}\over\partial \rho^{2}}+\tau_{2cl}
   {3\over\rho}{\partial\over\partial\rho}+
   {\partial^{2}\over\partial r^{2}}+
   {1\over r}{\partial\over\partial r}+{1\over r^{2}}
   {\partial^{2}\over\partial\theta^{2}}\Big]f_{1}\cr
   &\qquad=
   {3m_{1}m_{2}m_{3}m_{4}\tau_{2cl}\over \pi}
   {cos(4\theta)\over r^{4}}
   \Big(-{16N\pi\over \big[\rho^{2}+\tau_{2cl}r^{2}\big]^{3}}
   +{96N\pi \rho^{2}\over \big[\rho^{2}+\tau_{2cl}r^{2}\big]^{4}.}
   \Big).}} $$

\noindent
By inspection, it is clear that the angular dependence of $f_{1}$ is given
by $f_{1}=cos(4\theta)g(r,\rho).$ The function $g$ satisfies

\eqn\Forg
{\eqalign{\Big[\tau_{2}^{(0)}&
{\partial^{2}\over\partial \rho^{2}}+\tau^{(0)}_{2}
{3\over\rho}{\partial\over\partial\rho}+
{\partial^{2}\over\partial r^{2}}+
{1\over r}{\partial\over\partial r}-{16\over r^{2}}\Big]g\cr
&\qquad=
{3m_{1}m_{2}m_{3}m_{4}\tau_{0}^{(2)}\over \pi}
{1\over r^{4}}
\Big(-{16N\pi\over \big[\rho^{2}+\tau_{2}^{(0)}r^{2}\big]^{3}}
+{96N\pi \rho^{2}\over \big[\rho^{2}+\tau_{2}^{(0)}r^{2}\big]^{4}.}
\Big).}}

\noindent
This equation admits a power series solution. To set up the solution, 
notice 
that both $r$ and $\rho$ have the dimensions of length ($L$). It is not 
difficult to see that $g$ has dimension $L^{-8}$. Thus, by dimensional
analysis, it must have an expansion of the form

\eqn\Expnsn
{g=\sum_{m}{c_{m}\over r^{m}\rho^{8-m}}.}

\noindent
For consisteny, we require that $g\to 0$ at least as $r^{-4}$ as $ r\to\infty$.
If this is not the case, $f_{1}$ is not a small correction to $f_{0}$. 
Thus, we restrict $m\ge 4$ in \Expnsn. With this restriction, after inserting 
\Expnsn\ into \Forg, one finds for the first few $c_{m}$:

$${\eqalign{c_{m}&=0,\qquad m<8,\qquad
c_{8}=\alpha m_{1}m_{2}m_{3}m_{4}N,\cr
c_{9}&=0\qquad
c_{10}=-6m_{1}m_{2}m_{3}m_{4}N\Big[
{1+\alpha\over (\tau_{2cl})^{3}}
\Big].}}$$

\noindent
The full solution is not needed, since only $f_{1}$ at $y=0$ enters the
probe action. Notice that this solution for $f_{1}$ is labeled by an 
arbitrary parameter $\alpha$ which cannot be fixed by the above iterative calculation. It is an interesting
open question to see if $\alpha$ can be fixed by a more sophisticated
analysis \rUs. The correction to the leading term in $f$

$$ f(y=0,a,\bar{a})={4N\pi\over (\tau_{2cl}a\bar{a})^{2}}+
\alpha m_{1}m_{2}m_{3}m_{4}N{1\over 2(\bar{a}a)^{2}}
\Big({q\over a^{4}}+{\bar{q}\over\bar{a}^{4}}\Big)
+O\Big({1\over |a|^{12}}\Big)$$

\noindent
represents a correction to the AdS$_{5}$ geometry. This correction is
expected because we are no longer dealing with a conformal field theory.
Notice that the AdS$_{5}$ geometry is recovered in the limit of large energies
($|a|\to\infty$) and in the limit of massless matter $m_{i}\to 0$.
Expanding the probe action in this background, we find that the pure scalar
terms read

$$ {\eqalign{S={T_{3}\over 2}\int\Big(&\Big[\tau_{2cl}-
{3\tau_{2cl} \over 2\pi}m_{1}m_{2}m_{3}m_{4}\Big(
{q\over a^4}+{\bar{q}\over \bar{a}^4}\Big)\Big]
\partial_{m}a\partial^{m}\bar{a}\cr
+&\Big[{2N\pi\over (a\bar{a})^{2}}+
{\alpha Nm_{1}m_{2}m_{3}m_{4}(\tau_{2cl})^{2}\over4(a\bar{a})^{2}}
\Big({q\over a^{4}}+{\bar{q}\over\bar{a}^{4}}\Big)\Big]
\partial_{n}a\partial^{n}a
\partial_{m}\bar{a}\partial^{m}\bar{a}
\Big).}} $$

\noindent
Notice that the correction to the four derivative terms has the structure of
the one instanton corrections computed using field theory. We have not been
able to fix $\alpha$ with our asymptotic analysis, so that the coefficient
of this correction could not be checked. As reviewed above, instanton 
effects explicitly break the conformal symmetry of the field theory. The
breaking of the $SO(2,4)$ conformal symmetry in the field theory is reflected 
in the corrections to the AdS$_{5}$ geometry, which break the $SO(2,4)$
isometry of the AdS$_{5}$ space. Note that the coupling runs with a power law.
Solutions of type-0 string theories with a power law running for the coupling
have been studied in \rAlv. Power law running of the coupling has also
played a prominent role in gauge-coupling unification in theories with
large internal dimensions\rDie. The presence of the large internal dimensions
is reflected in the fact that massive Kaluza-Klein modes run in loops of the
four dimensional theory. This gives rise to a power law running of the 
couplings. In \rGuby\ it was suggested that this effect may be responsible 
for the power law running of couplings in the IIB background discussed in 
that study. In our case, there is no need for effects due to large internal 
dimensions and the power law running of the coupling is simply explained
by instanton effects in the four dimensional field theory.

Before leaving this section we would like to make some comments on the 
supergravity interpretation of the leading $|a|^{-4}$ behaviour of the
four derivative terms. The operator on the worldvolume which couples to the
dilaton is given by\rDas\

$$ {\cal O}_{\phi}=-{1\over 4}F^{\mu\nu}F_{\mu\nu}. $$

\noindent
By expanding the Born-Infeld action one finds a four derivative term in 
the effective potential that has the form

$$ {\cal O}_{\phi}{\cal O}_{\phi}\Big[{2N\pi\over (a\bar{a})^{2}}+
{\alpha Nm_{1}m_{2}m_{3}m_{4}\tau_{2cl}^{2}\over 4(a\bar{a})^{2}}
\Big({q\over a^{4}}+{\bar{q}\over\bar{a}^{4}}\Big)\Big]. $$

\noindent
The leading term of $|a|^{-4}$ comes from the static massless propagator in 
the six dimensional transverse space. This term is due to exchange of a dilaton
and appears because the supergravity modes which couple to constant gauge fields
on the brane have zero momentum along the brane \rDas. 

\subsec{Instanton Effects and the static Quark-Antiquark Potential from
Supergravity}

In this section we study the static quark-antiquark potential, in 
the large $|\varphi|$ region where the background geometry described above
is valid. This corresponds to studing the effects of instantons on the 
static quark-antiquark potential in the large $N$ field theory. There is a large body
of evidence from lattice calculations that indicate that instanton effects
play a major role in the physics of light hadrons\rNeg. Below we will argue 
that the supergravity description provides a powerful new approach to these 
questions.

The energy of a quark-antiquark pair can be read off of the expectation value
of a Wilson loop. This Wilson loop is identified with a fundamental string
ending on the boundary of the asymptotically AdS$_{5}$ space\rLoop. The Wilson
loop configuration is thus obtained by minimizing the Nambu-Goto 
action\foot{In this expression $\alpha,\beta=\tau,\sigma,$ 
and $M,N=0,1,...,9$}

$$S={1\over 2\pi}\int d\tau d\sigma\int\sqrt{det(g_{MN}
\partial_{\alpha}x^{M}\partial_{\beta}x^{N})} $$

\noindent
The metric $g_{MN}$ felt by the strings is not the Einstein metric \frth, 
but rather the string frame metric. We are interested in a static string
configuration and take $\sigma=x^{1}$ and $\tau=x^{0}$. The string is at
a fixed $x^{2},x^{3},x^{4},x^{5},x^{6},x^{7}$ and $\theta$ where $\theta$
is the angular variable in the $(8,9)$ plane. In terms of the variable
$r^{2}\equiv a\bar{a},$ the Nambu-Goto action takes the
form

$$ S={T\over 2\pi}\int d\sigma
\sqrt{(\partial_{\sigma}r)^{2}+{1\over f\tau_{2}}}
={T\over 2\pi}\int d\sigma
\sqrt{(\partial_{\sigma}r)^{2}+{r^{4}\over a}-{b\over a^{2}}} $$

$$a={4N\pi\over \tau_{2cl}},\quad
  b=\cos (4\theta)m_{1}m_{2}m_{3}m_{4}N\Big(
{\alpha\tau_{2cl}\over 2}-{6\over \tau_{2cl}}\Big) $$

\noindent
where $T=\int d\tau$. We have dropped terms of $O(r^{-4})$ in the square 
root above. The solution to the Euler-Lagrange equations of motion following 
from this action is obtained in the usual way: The action does not
depend explicitly on $\sigma$ so that the Hamiltonian in the $\sigma$
direction is a constant of the motion

$$
{{r^{4}\over a}-{b\over a^{2}}\over
\sqrt{(\partial_{\sigma}r)^{2}+{r^{4}\over a}-{b\over a^{2}}}}=
const=\sqrt{{r_{0}^{4}\over a}-{b\over a^{2}}}
$$

\noindent
where $r_{0}$ is the minimal value of $r$. By symmetry we have
$r(\sigma=0)=r_{0}$. It is now straight forward to obtain

$$
\sigma=\sqrt{{r_{0}^{4}\over a}-{b\over a^{2}}}
\int_{r_{0}}^{r} {dy\over
\sqrt{({y^{4}\over a}-{b\over a^{2}})
      ({y^{4}\over a}+{r_{0}^{4}\over a})} }.$$

\noindent
The string endpoints are at the boundary of the asymptotically AdS$_{5}$
space ($r\to\infty$) so that we can trade the integration constant $r_{0}$
for the distance $L$ between the quark and anti-quark

$$
L=2\sqrt{{r_{0}^{4}\over a}-{b\over a^{2}}}
\int_{r_{0}}^{\infty} {dy\over
\sqrt{({y^{4}\over a}-{b\over a^{2}})
      ({y^{4}\over a}+{r_{0}^{4}\over a})} }.$$

\noindent
The energy is now computed by evaluating our action at this classical 
solution. After subtracting twice the self-energy of a quark, we obtain 
the following result for the quark-antiquark potential

$$
E={1\over\pi\sqrt{({r_{0}^{4}\over a}-{b\over a^{2}})}}
\int_{r_{0}}^{\infty}dy\left(
{\sqrt{(y^{4}-{b\over a})}
\over\sqrt{y^{4}+r_{0}^{4}}}-1\right).
$$ 

\noindent
To extract the dependence of this energy on the quark-antiquark separation $L$
we need to determine $r_{0}$ as a function of $L$. The expression for the 
energy given above is identical to the static potential for the 
quark-antiquark pair in the ${\cal N}=4$ theory at finite temperature\rSon.
>From the results of \rSon\ we know that $E(L)$ has the form

$$
E=-{c_{1}\over L}-c_{2}L^{3}.
$$

\noindent
The constant $c_{1}$ is positive. In the limit that $m_{i}\to 0$, $c_{2}\to 0$
and we regain the ${1\over L}$ dependence, a fact which is determined by
conformal invariance. The sign of the constant $c_{2}$ is dependent on 
$\theta.$ For $c_{2}$ positive (negative) we have a screening (antiscreening)
of the quark-antiquark pair due to the instantons.
This expression can't be trusted for very large $L$: for larger
and larger $L$ the Wilson loop is able to move further and further into the
bulk. Our solution is however only valid for large $r$, so that the Wilson loop
begins to explore regions in the bulk for which our solution is not valid. The
long distance behaviour of the quark-antiquark potential could be extracted 
from the exact supergravity background.

\newsec{Pure Gauge ${\cal N}=2$ Super Yang-Mills Theory with Gauge Group 
$SU(2)$}

In this section we obtain the leading correction to the four derivative terms
in both the field theory and the supergravity descriptions.

\subsec{Field Theory Results}

In the case where there are no flavor multiplets, the perturbative beta 
function does not vanish and the field theory is asymptotically free. We 
will focus attention on the perturbative contributions to the two and four 
derivative terms appearing in the low energy effective action. The one loop
results for the K\"ahler metric and real function ${\cal H}$ are

$$ K_{\varphi\bar{\varphi}}\sim 
{\log (\varphi\bar{\varphi}/\Lambda^{2})\over\varphi\bar{\varphi}}, $$

$$ {\cal H}(A,\bar{A})\sim \log \Big({A\over\Lambda}\Big)
\log\Big({\bar{A}\over\Lambda}\Big). $$

\noindent
This leads to the following four derivative term for the scalars, after
performing the field redefinition\rHo\

$$ \eqalign{S&=\int d^{4}x 
   (\partial^{m}\varphi\partial_{m}\varphi)
   (\partial^{n}\bar{\varphi}\partial_{n}\bar{\varphi})
   {8+4\log\Big({\varphi\bar{\varphi}\over\Lambda^{2}}\Big)
   +\Big[\log\Big({\varphi\bar{\varphi}\over\Lambda^{2}}\Big)\Big]^{2}\over
   \varphi^{2}\bar{\varphi}^{2}
   \Big[\log\Big({\varphi\bar{\varphi}\over\Lambda^{2}}\Big)\Big]^{2}}\Big]\cr
  &=\int d^{4}x (\partial^{m}\varphi\partial_{m}\varphi)
  (\partial^{n}\bar{\varphi}\partial_{n}\bar{\varphi})
  \Big[{1\over (\varphi\bar{\varphi})^{2}}+
  O\Big({1\over (\varphi\bar{\varphi})^{2}log|\varphi|}\Big)\Big].}$$

\noindent
The $|\varphi|^{-4}$ fall off at large separations (large $|\varphi|$) 
suggests that the dominant interaction between the branes is again due to 
the exchange of massless supergravity modes propagating in the six 
dimensional space transverse to the three branes.

\subsec{Supergravity Results}

The problem of finding the relevant background geometry corresponding to
the asymptotically free gauge theory is considerably more complicated. The
Laplace equation \sxh\ becomes (we will show all $N$ dependence in this 
section)

$$ \Big[{N\over\lambda }
\Big({8\pi\over g_{cl}^{2}}+{6\over\pi}+{2\over\pi}
\log \Big({a\bar{a}\over\Lambda^{2}}\Big)
\Big){\partial^{2}\over\partial y^{2}}
+{\partial\over\partial a}{\partial\over\partial \bar{a}}\Big]f=
-(2\pi )^{4}N\delta^{(4)}(y)\delta^{(2)}(a). $$

\noindent
Performing a Fourier transform on the $y$ variables and working in the large 
$|a|$ region (which corresponds to the semi-classical regime of the field 
theory) we find

$$ \Big[{\partial\over\partial a}{\partial\over\partial \bar{a}}
-k^{2}{N\over\lambda}
\Big({8\pi\over g_{cl}^{2}}+{6\over\pi}+{2\over\pi}
\log \Big({a\bar{a}\over\Lambda^{2}}\Big)\Big)
\Big]f=0. $$

\noindent
We have not been able to solve this equation exactly. However an approximate
solution in the large $|a|$ region is given by

$$ f\approx Ne^{-{2k\sqrt{N}|a|\over\sqrt{\lambda\pi}}
\sqrt{{2\pi^{2}\over g_{cl}^{2}}+{1\over 2}+\log(|a|)}},$$

\noindent
Using this approximate solution we obtain

$$ f(y=0,a,\bar{a})=\int d^{4}k f(k,a,\bar{a})=
{12\pi^{2}\lambda^{2}\over 16N(a\bar{a})^{2}
\Big(\log\Big[{\sqrt{a\bar{a}}\over\Lambda}exp({2\pi^{2}\over g_{cl}^{2}}+{1\over 2})\Big]
\Big)^{2}}. $$

\noindent
This result determines the coefficient of the four derivative terms

$$
\tau_{2}^{2}f(y=0,a,\bar{a})\sim {N\over (a\bar{a})^{2}}+
O\Big({1\over (a\bar{a})^{2}log|a|}\Big). $$

\noindent
This is exactly the same behaviour as obtained from the field theory
analysis.

The complex coupling $\tau$ in this supergravity background has a logarithmic
dependence on $a\bar{a}$ corresponding to the logarithmic dependence of the 
field theory coupling on the energy scale. Gravity solutions that have
couplings with this logarithmic dependence have been constructed in 
type-0 theories\rIgK. 

We should now address the validity of this computation. There are two potential
sources of corrections to the supergravity background - string loop effects
and curvature corrections. At large $N$ and large 't Hooft coupling both of
these types of corrections are small and supergravity is a reliable
description of the background. As the 't Hooft coupling decreases, curvature 
corrections become
important and the uncorrected supergravity can no longer be trusted\rWK. The
uncorrected supergravity {\it does not} correctly describe the large $N$ 
{\it perturbative} field theory. We would like to determine wether the 
uncorrected supergravity is a valid description for the 
perturbative field theory living 
on the probe, which is a different question. The simplest way to asses the 
validity of the supergravity description is simply to compute the square of
the Ricci tensor. This calculation should be carried 
out in string frame because
we are interested in the region in which the dilaton is going to zero. We will
show both the Einstein and string frame results in what follows. In the large
$|a|$ region the Einstein frame metric that we have computed above takes the
form\foot{In what follows $\mu,\nu=0,1,2,3$; $i,j=4,5,6,7$ and
$M,N=0,1,...,9$.}

$$ g_{MN}^{(e)}dx^{M}dx^{N}={\sqrt{N}\over\lambda}\Big(a\bar{a}\log|a|
\eta_{\mu\nu}dx^{\mu}dx^{\nu}
+{1\over a\bar{a}\log |a|}\Big(\delta_{ij}dx^{\prime i}dx^{\prime j}
+\lambda\log|a| dad\bar{a}\Big)\Big),$$

\noindent
where we have rescaled $x^{i}\to x^{\prime i}$ where $x'=x\lambda/\sqrt{N}$. 
The corresponding string frame metric is

$$ g_{MN}^{(s)}dx^{M}dx^{N}={1\over\sqrt{\lambda}}
\Big(a\bar{a}(\log|a|)^{1/2}
\eta_{\mu\nu}dx^{\mu}dx^{\nu}
+{1\over a\bar{a}(\log |a|)^{3/2}}\Big(\delta_{ij}dx^{\prime i}dx^{\prime j}
+\log|a|\lambda dad\bar{a}\Big)\Big).$$

\noindent
The leading contribution to the square of the Ricci tensor in the Einstein
frame is a constant

$$R_{MN}R^{MN}={32\over N}.$$

\noindent
In the string frame, the leading contribution to the square of the Ricci tensor
diverges logarithmically for large $|a|$

$$R_{MN}R^{MN}={32\over \lambda}\log |a|.$$

\noindent
To interpret these results, note that the Yang-Mills coupling squared is
$g^{2}=\lambda(N\log|a|)^{-1}$, so that the 't Hooft coupling is 
$\lambda_{T}\equiv g^{2}N=\lambda/\log|a|.$ We see that the square of the Ricci 
tensor in the string frame is inversely proportional to the 't Hooft coupling,
so that we recover the well known result that curvature effects in the background 
are small at large $\lambda_{T}$. The perturbative probe field theory is valid for
 $|a|>>1$.
It is clear that in the $N\to\infty$ limit, large $\lambda_{T}$ and large $|a|$ 
are compatible, i.e. in the large $N$ limit, the 't Hooft coupling is large even
when the probe worldvolume field theory is perturbative. Moving
to smaller $|a|$ one would need to correct the asymptotic solution for the background 
that we have
found. The supergravity solution in this region captures the strong
coupling dynamics of the asymptotically free gauge theory on the probe.

Finally, we note that the effects that we have computed in this section are 
linear in both the number of source branes and the number of probe branes. The
supergravity will not capture effects which do not have this linear dependence.

{\it Acknowledgements:} It is a pleasure to thank Zach Guralnik,Antal Jevicki, 
David Lowe
and Sanjaye Ramgoolam for useful discussions and Jo\~ao Nunes and Martin
Rocek for helpful email correspondence. The work of RdMK is supported by
a South African FRD bursary. The work of APC and JPR is partially 
supported by the FRD under grant number GUN-2034479.

\listrefs
\vfill\eject
\bye